\documentclass[conference]{IEEEtran}
\IEEEoverridecommandlockouts
\usepackage{cite}
\usepackage{amsmath,amssymb,amsfonts}
\usepackage{algorithmic}
\usepackage{graphicx}
\usepackage{textcomp}
\usepackage{xcolor}
\usepackage{bm}
\usepackage{mathtools}
\usepackage{multirow}
\usepackage{pgfplots}

\pgfplotsset{compat=1.11}

\def\BibTeX{{\rm B\kern-.05em{\sc i\kern-.025em b}\kern-.08em
    T\kern-.1667em\lower.7ex\hbox{E}\kern-.125emX}}

\begin{document}

\title{Algorithms for Piecewise Constant Signal 
Approximations\thanks{This project has received funding from the European 
Research Council (ERC) under the European Union's Horizon 2020 research 
and innovation programme (grant agreement No 741215).
It has been initiated at a joint visit of Alfred Bruckstein and Joachim
Weickert to the Isaac Newton Institute, which was supported by
EPSRC Grant Number EP/K032208/1 and by a Rothschild Distinguished
Visiting Fellowship.
}}

\author{%
%\IEEEauthorblockN{1\textsuperscript{st} Leif Bergerhoff}
\IEEEauthorblockN{Leif Bergerhoff}
\IEEEauthorblockA{\textit{Mathematical Image Analysis Group}\\
\textit{Saarland University}\\
66041 Saarbr\"ucken, Germany\\
bergerhoff@mia.uni-saarland.de}
\and
%\IEEEauthorblockN{2\textsuperscript{nd} Joachim Weickert}
\IEEEauthorblockN{Joachim Weickert}
\IEEEauthorblockA{\textit{Mathematical Image Analysis Group}\\
\textit{Saarland University}\\
66041 Saarbr\"ucken, Germany\\
weickert@mia.uni-saarland.de}
\and
%\IEEEauthorblockN{3\textsuperscript{rd} Yehuda Dar}
\IEEEauthorblockN{Yehuda Dar}
\IEEEauthorblockA{\textit{Computer Science Department} \\
\textit{Technion -- Israel Institute of Technology}\\
Haifa 3200003, Israel\\
ydar@cs.technion.ac.il}
}

\maketitle

\begin{abstract}
We consider the problem of finding optimal piecewise constant approximations 
of one-dimensional signals. These approximations should consist of a 
specified number of segments (samples) and minimise the mean squared error 
to the original signal. We formalise this goal as a discrete nonconvex 
optimisation problem, for which we study two algorithms. First we 
reformulate a recent adaptive sampling method by Dar and Bruckstein 
in a compact and transparent way. This allows us to analyse its limitations 
when it comes to violations of its three key assumptions: signal smoothness, 
local linearity, and error balancing. 
As a remedy, we propose a direct optimisation approach which does not
rely on any of these assumptions and employs a particle swarm optimisation 
algorithm. Our experiments show that for nonsmooth signals or low sample 
numbers, the direct optimisation approach offers substantial qualitative 
advantages over the Dar--Bruckstein method.
As a more general contribution, we disprove the optimality of the principle
of error balancing for optimising data in the $\ell^2$ norm.
\end{abstract}


%%%%%%%%%%%%%%%%%%%%%%%%%%%%%%%%%%%%%%%%%%%%%%%%%%%%%%%%%%%%%%%%%%%%%%%%%%%

\section{Introduction}

One of the essential concepts in signal processing is the sampling 
and reconstruction of continuous signals. 
The classical sampling theory (see e.g. \cite{Je77} for a review) 
considers uniform sampling and provides conditions under which the 
sampling allows lossless signal reconstruction. These concepts have 
been extended to nonuniform sampling where one can adapt the sampling 
rate to the local signal bandwidth \cite{Ho68, CPL85, BPP98, WO07}.
% A common strategy to obtain a nonuniform sampling is to apply a nonlinear 
% transform to the time axis of the signal first, followed by a uniform 
% sampling in transformation space afterwards \cite{CPL85, BPP98, WO07}.

If one aims at lossy signal representations to achieve higher
compression rates, other options can be preferable. Recently Dar and 
Bruckstein \cite{DB16} have introduced a simple and efficient adaptive 
sampling strategy for approximating 1-D signals by piecewise constant 
functions. 
It involves three assumptions: smoothness, local linearity, and error 
balancing. In practice, however, signals can be nonsmooth, they can 
violate local linearity, and the optimality of error balancing is unclear. 
Thus, finding the optimal approach for the general case remains an open 
problem.

%-------------------------------------------------------------------------
 
\subsection{Our Contributions}

To address this problem, we first revisit the Dar--Bruckstein model 
by deriving it in a simpler, alternative way which is inspired by 
the work of Belhachmi et al. \cite{BBBW09}. Its analysis enables us 
to quantify the effects of violating local linearity and to disprove
the optimality of error balancing. As a remedy, we propose an energy
minimisation model that does not rely on smoothness, local linearity, 
or error balancing. It favours globally optimal piecewise constant 
signal approximations which minimise the mean squared error (MSE). 
This requires to solve a nonconvex optimisation problem, for which a 
particle swarm optimisation (PSO) algorithm \cite{KE95} performs 
well. Experiments show that the quality of our novel approach
can exceed the one of the Dar--Bruckstein method.

%-------------------------------------------------------------------------

\subsection{Structure of the Paper}

In Section \ref{sec:theory} we introduce the underlying approximation 
problem, reformulate and analyse the Dar--Bruckstein model, and we 
propose our novel method along with the PSO algorithm.  
Section \ref{sec:experiments} presents experimental comparisons of both
approaches with a smooth, a nonsmooth, and a noisy signal. We conclude
with a summary and an outlook in Section \ref{sec:summary}.

%%%%%%%%%%%%%%%%%%%%%%%%%%%%%%%%%%%%%%%%%%%%%%%%%%%%%%%%%%%%%%%%%%%%%%%%%%%

\section{Modelling}
\label{sec:theory}

In this section we formalise our approximation problem and discuss two 
different solution strategies: 1. The Dar--Bruckstein approach for which 
we give a new and compact derivation and an analysis of its limitations. 
2. Our novel direct optimisation method which aims at overcoming these 
limitations by renouncing assumptions of the Dar--Bruckstein model 
that may be violated or cause suboptimal solutions.

%---------------------------------------------------------------------------

\subsection{Problem Statement}
\label{sec:theory_problem}

Consider a signal domain $[a, b] \subset \mathbb{R}$ and some 
integrable 1-D signal $f: [a,b] \to \mathbb{R}$. We want 
to approximate $f$ by a piecewise constant signal $u: [a,b] \to \mathbb{R}$
that has $N$ segments (samples) and minimises the MSE w.r.t.~$f$.
This requires to find $N\!-\!1$ segment boundaries $x_1$, $x_2$, ..., 
$x_{N-1}$ with
\begin{equation}
 a < x_1 < x_2 < \ldots < x_{N-1} < b.
\end{equation}
It is convenient to introduce also $x_0:=a$ and $x_N:=b$.
Then we approximate $f$ by a piecewise constant signal of type
\begin{equation}
\label{eq:u}
 u(x) :=
 \begin{cases}
 u_i, & \text{for } x \in [x_i, x_{i+1}) \text{ and } i<N\!-\!1,\\
 u_{N-1}, & \text{for } x \in [x_{N-1}, x_N].\\
\end{cases}
\end{equation}
Since we aim at an $\ell^2$-optimal approximation, each constant $u_i$ is
given by the mean value of $f$ in $[x_i,x_{i+1}]$:
\begin{equation}
 u_i \;:=\;
 \frac{1}{x_{i+1}\!-\! x_i} \,
 \int \limits_{x_i}^{\mathclap{x_{i+1}}} f(y) \,\, dy. 
\end{equation}
We observe that $u$ is completely determined by $f$ and the segment boundary 
vector $\bm{x} := (x_1, x_2,\ldots,x_{N-1})^\top$. Thus, our problem 
comes down to minimising the discrete energy
\begin{equation}
\label{eq:energy}
 E(\bm{x}) \;=\;
 \frac{1}{b\!-\!a} \, \sum \limits_{i = 0}^{N-1} \,
 \int \limits_{x_i}^{\mathclap{x_{i+1}}} \big(
 f(y) - u_i \big)^{2} \,\, dy \,.
\end{equation}
Depending on the application, this may be interpreted as function 
approximation, adaptive sampling, segmentation, or lossy signal 
compression.\\
Although the energy (\ref{eq:energy}) does not look very complicated, in 
general it is nonsmooth, nonconvex, and may have many local minima. This 
is illustrated in Fig. \ref{fig:example_signal} for a piecewise constant 
signal $f(x)$, for which we want to find an approximation $u(x)$ with a 
single jump position $x_1$. This corresponds to the simplest nontrivial 
scenario $N=2$.

%...........................................................................

\begin{figure}
    \begin{center}
        \begin{tikzpicture}
        \begin{axis}[%
        xmin=0, xmax=4,
        width=0.42\textwidth,
        xlabel={$x$},
        ylabel={Signal Value}]
        \addplot[black, const plot, thick]
        coordinates {(0, 8) (1, 5.5) (2, 2) (3, 3) (4, 3)};
        \addplot[red, const plot, thick]
        coordinates {(0, 6.75) (2, 2.5) (4, 2.5)};
        \legend{$f(x)$, $u(x)$}
        \end{axis}
%        \end{tikzpicture}
        \end{tikzpicture}\\[1em]
        \begin{tikzpicture}
        \begin{axis}[%
        xmin=0, xmax=4,
        width=0.42\textwidth,
        xlabel={Jump Position $x_1$},
        ylabel={Energy $E(x_1)$}]
        \addplot[red, thick] table[x={x}, y={fx}] {data/sample_mse.dat};
        \end{axis}
        \end{tikzpicture}
    \end{center}
    \caption{%
        \textbf{(a) Top:}
        A piecewise constant signal $f(x)$. Its optimal piecewise constant 
        approximation $u(x)$ with $N = 2$ segments is obtained for 
        $x_1 := 2$. 
        \textbf{(b) Bottom:} The corresponding energy / MSE curve as a 
        function of $x_1$ is nonsmooth and nonconvex, and it has two 
        local minima.}
    \label{fig:example_signal}
\end{figure}

%-------------------------------------------------------------------------------

\subsection{Compact Reformulation of the Dar--Bruckstein Method}
\label{sec:theory_dar}

Recently Dar and Bruckstein \cite{DB16} have proposed an approach to 
solve the problem in Section \ref{sec:theory_problem} very efficiently. 
To explain the underlying ideas and assumptions in a simple and 
transparent way, we reformulate its derivation. This reformulation is 
inspired by work of Belhachmi et al.~\cite[Section 6]{BBBW09}.

Denoting the squared error in the interval $[x_i, x_{i+1}]$ by
\begin{equation}
 \label{eq:err_01}
 e_i \;:=\;
 \int \limits_{x_i}^{x_{i+1}} \big(
 f(y) - u_i
 \big)^{2} \,\, dy\,,
\end{equation}
we can write the energy function (\ref{eq:energy}) as
\begin{equation}
\label{eq:energy2}
 E(\bm{x}) \;=\; \frac{1}{b - a} \, \sum \limits_{i = 0}^{N-1} e_i\,.
\end{equation}

Dar and Bruckstein assume that the input signal $f$ is a continuously 
differentiable ($C^1$) function and that $N$ is large enough such that $f$ 
can be approximated well by a linear function within each interval 
$[x_i, x_{i+1}]$ for $i = 0, 1, \ldots, N\!-\!1$. Thus, in $(x_i, x_{i+1})$  
we have 
\begin{align}
% \label{eq:u} 
 f'(x) &\;=\; \frac{f(x_{i+1}) - f(x_i)}{x_{i+1}-x_i} \;=:\; f_i'\,,\\
 f(x)  &\;=\; f(x_i) \,+\, (x-x_i) \, f_i'\,, \label{eq:f}\\ 
 u_i   &\;=\; \tfrac{1}{2} \, (f(x_i) + f(x_{i+1}))\,.
% \label{eq:u}
\end{align}
Using this in (\ref{eq:err_01}) and applying some simple calculations yields 
\begin{equation}
 \label{eq:errordb}
 e_i \;=\; \tfrac{1}{12} \, h_i^3 \, f_i'^{\,2}
\end{equation}
where $\,h_i := x_{i+1}-x_i\,$ denotes the interval width. 

As a heuristics for minimising the global energy (\ref{eq:energy2}), one 
may assume that $\bm{x}$ is optimal if all local errors $e_i$ are balanced.
Using $e_0=e_1=...=e_{N-1}=\mbox{const.}$ with (\ref{eq:errordb}) gives 
the following proportionalities: 
\begin{equation}
 f_i'^{\,2} \;\sim\; \frac{1}{h_i^3} \quad \Longrightarrow \quad
 \frac{1}{h_i} \;\sim\; \sqrt[3] {f_i'^{\,2}}\: .
\end{equation}
Since $1/h_i$ can be seen as a measure for the local density of 
the sampling points, one should choose the interval 
boundaries for optimal sampling proportional to
$\sqrt[3]{f_i'^{\,2}}$.
Consequently, Dar and Bruckstein select $\bm{x}$ such that 
every segment $[x_i, x_{i+1}]$ contains the same amount of the 
cube root of the squared signal derivative. More precisely:
\begin{equation}
\label{eq:yehuda_sol}
\int \limits_{\mathclap{x_{i}}}^{x_{i+1}} \! \sqrt[3]{\big(f'(y)\big)^2} \, dy 
\;=\;
\frac1N \, \int \limits_a^b \! \sqrt[3]{\big(f'(y)\big)^2} \, dy 
\;=:\; T_{opt} 
\end{equation}
for $i = 0, 1, \ldots, N\!-\!1$.
The threshold $T_{opt}$ is computed a priori.  
Thus, the analytical formula \eqref{eq:yehuda_sol} allows to estimate the 
interval boundaries $\bm{x}$ in a simple and efficient way.

%--------------------------------------------------------------------------

\subsection{Limitations of the Dar--Bruckstein Method}

We have seen that the Dar--Bruckstein approach relies on three assumptions: 
$C^1$-smoothness, local linearity, and error balancing. Let us now analyse 
the impact of these assumptions on the optimality of the method in detail. 
\begin{itemize}
\item Obviously the smoothness assumption on $f$ is violated if
      the signal is nondifferentiable or noisy.
\item To quantify inaccuracies caused by violations of the local linearity 
      assumption, we derive a formula for $e_i$ that does not use this 
      assumption. We can rewrite \eqref{eq:err_01} as
      \begin{equation}
       \label{eq:yehuda_01}
       e_i =
       \int \limits_{x_i}^{\mathclap{x_{i+1}}} \big( f(y) - f(\xi_i) \big)^2
       \,\, dy\,,
      \end{equation}
      where we have used the continuity of $f$, which guarantees that 
      there exists a $\xi_i \in [x_i, x_{i+1}]$ with $f(\xi_i) = u_i$.
      With the mean value theorem, Equation \eqref{eq:yehuda_01} becomes 
      \begin{align}
       e_i &\,=\, \int \limits_{x_i}^{\mathclap{x_{i+1}}}
               (\xi_i - y)^2 \, \big( f'(\theta_i) \big)^2 \, dy\\
           &\,=\, \tfrac{1}{3}\,\left((x_{i+1}\!-\!\xi_i)^3 + 
               (\xi_i\!-\!x_i)^3 \right) \, \big( f'(\theta_i) \big)^2
      \label{eq:yehuda_02}
      \end{align}
      for a suitable $\theta_i \in [x_i, x_{i+1}]$.
      Using $\,h_i = x_{i+1}\!-\!x_i\,$ and defining 
      $\,\eta_i:=(x_{i+1}\!-\!\xi_i) / h_i\,$
      allows to rewrite (\ref{eq:yehuda_02}) as 
      \begin{equation}
       \label{eq:exact}
       e_i \;=\; \tfrac{1}{3} \left(1-3\eta_i+3\eta_i^2\right) \, h_i^3\:
       \big( f'(\theta_i) \big)^2\,.
      \end{equation}
      Comparing the exact error (\ref{eq:exact}) with the error 
      (\ref{eq:errordb}) that exploits local linearity shows the following:
      Since $\xi_i \in [x_i, x_{i+1}]$, we know that $\eta_i \in [0,1]$.
      However, only for $\eta_i=\tfrac{1}{2}$, we obtain 
      $\tfrac{1}{3} \left(1-3\eta_i+3\eta_i^2\right)=\tfrac{1}{12}$. 
      It the worst case with $\eta_i=0$ or $1$, this factor becomes 
      $\tfrac{1}{3}$.
      Moreover, since $f \in C^1[a,b]$, there exist constants 
      $\,m:=\min_{[a,b]} f'\,$ and 
      $\,M:=\max_{[a,b]} f'\,$. 
      Thus, $\big( f'(\theta_i) \big)^2$ can attain any value between 
      $m^2$ and $M^2$, which can differ substantially from $f_i'^2$. 
      This shows that without local linearity, $(\ref{eq:errordb})$ can be 
      violated severely.
      Moreover, (\ref{eq:yehuda_sol}) does no longer balance 
      the errors then.
\item While the principle of error balancing sounds plausible, one cannot 
      prove that it is fulfilled for the globally optimal $u$ which 
      minimises the MSE. Actually, already Fig.~\ref{fig:example_signal}(a)
      serves as counterexample: The error in the left segment of $u$ is 
      clearly larger than in the right segment.  
\end{itemize}

%-------------------------------------------------------------------------------

\subsection{A Novel Direct Optimisation Approach}
\label{sec:theory_model}

The preceding discussion shows that it can be desirable to renounce
all three assumptions of the Dar--Bruckstein model. Interestingly, 
there is a surprisingly simple solution: We can rely directly on
the discrete model (\ref{eq:energy}), which is perfect from a
modelling viewpoint. However, we have to deal with a challenging 
nonsmooth and nonconvex optimisation problem with numerous local 
minima. 
Since we cannot expect to find an efficient algorithm with formal 
convergence guarantees to a global minimum, we use a nature-inspired 
metaheuristic that ends up in a good local minimum. Based on our tests, 
we recommend to minimise (\ref{eq:energy}) by a Particle Swarm 
Optimisation (PSO) approach. 
PSO is an iterative global optimisation technique for nonlinear 
functions. It emerged from a simulation of social behaviour \cite{KE95}. 
The term ``swarm'' refers to a specified number of $n$ virtual particles 
which explore the solution space while interacting among each other.
The $i$-th particle represents one solution $\bm{x}_i$ with corresponding 
energy $E(\bm{x}_i)$.

In our work we apply the Standard Particle Swarm Optimisation 2011 
algorithm (SPSO-2011), which uses an adaptive random particle 
neighbourhood topology and features rotation invariance 
\cite{ZBCR13}. 
The initial particle positions $\bm{x}_i^0$ with $i = 1, 2, \ldots, n$
are uniformly distributed in the solution space.
In iteration step $k$, the particle position $\bm{x}_i^k$ 
is updated as follows: 
\begin{align}
 \label{eq:pso1} 
 \bm{p}_i^k & \;=\; \bm{x}_i^k \,+\, c_1 \, \bm{u}_1^k \odot 
                \left(\bm{P}_i^k \!-\! \bm{x}_i^k\right),\\
 \label{eq:pso2} 
 \bm{l}_i^k & \;=\; \bm{x}_i^k \,+\, c_2 \, \bm{u}_2^k \odot 
                \left(\bm{L}_i^k \!-\! \bm{x}_i^k\right),\\
 \label{eq:pso3} 
 \bm{g}_i^k & \;=\; \tfrac{1}{3} 
                \left(\bm{x}_i^k + \bm{p}_i^k + \bm{l}_i^k\right),\\
 \label{eq:pso4} 
 \bm{v}_i^{k+1} & \;=\; \omega \, \bm{v}_i^k 
                \,+\, \mathcal{H}_i \left(\bm{g}_i^k, 
                \, |\bm{g}_i^k\!-\!\bm{x}_i^k|\right) \,-\, \bm{x}_i^k,\\
 \label{eq:pso5} 
 \bm{x}_i^{k+1} & \;=\; \bm{x}_i^k \,+\, \bm{v}_i^{k+1},
\end{align}
where $\bm{P}_i^k$ denotes the previously best position of particle $i$ 
w.r.t.~the energy $E$, and
$\bm{L}_i^k$ is the best previous position
%of its $K$
%nearest
%neighbours.
%among itself and its $K$ neighbours.
within its neighbourhood of size $K$.
The symbol $\odot$ denotes element-wise vector multiplication, 
and $\bm{u}_1^k$ and 
$\bm{u}_2^k$ are independent and uniformly distributed random vectors with
components in $[0, 1]$. The scalars $c_1$, $c_2$, and $\omega$ 
are nonnegative weights.
$\mathcal{H}_i(\bm{g}_i^k, \, |\bm{g}_i^k-\bm{x}_i^k|)$ selects a random 
point from the 
% $(N\!+\!1)$--dimensional 
hypersphere around $\bm{g}_i^k$ with radius $|\bm{g}_i^k\!-\! \bm{x}_i^k|$ 
in the Euclidean norm $|\,.\,|$.

This algorithm can be understood as follows.
Equation (\ref{eq:pso1}) incorporates the successful history of particle 
$i$ by defining a point $\bm{p}_i^k$ near $\bm{P}_i^k$. 
Equation (\ref{eq:pso2}) expresses the knowledge of its neighbours by 
specifying a point $\bm{l}_i^k$ near $\bm{L}_i^k$.
In (\ref{eq:pso3}) we compute the centre of gravity 
$\bm{g}_i^k$ of $\bm{x}_i^k$, $\bm{p}_i^k$, and $\bm{l}_i^k$.
Equation (\ref{eq:pso4}) involves $\bm{g}_i^k$ to update the velocity 
$\bm{v}_i^{k+1}$ of particle $i$.
This velocity is used in (\ref{eq:pso5}) to move $\bm{x}_i^k$ to its new 
position $\bm{x}_i^{k+1}$.
If the global optimum shows no improvement then each particle randomly selects 
$K$ new neighbours.
%The neighbourhood composition of each particle
%size $K$
%changes randomly
The algorithm stops if a maximum number of iterations or 
a tolerable energy threshold is reached.
For more details, we refer to \cite{ZBCR13}.
% While the optimality of the Dar--Bruckstein method is compromised by two 
% modelling aspects (smoothness, local linearity) and one optimisation issue 
% (error balancing), the direct optimisation only suffers from its  
% PSO suboptimality.

%%%%%%%%%%%%%%%%%%%%%%%%%%%%%%%%%%%%%%%%%%%%%%%%%%%%%%%%%%%%%%%%%%%%%%%%%%%

%\clearpage
\section{Experiments}
\label{sec:experiments}

Let us now evaluate the approximation quality of the Dar--Bruckstein method
and our direct optimisation approach.

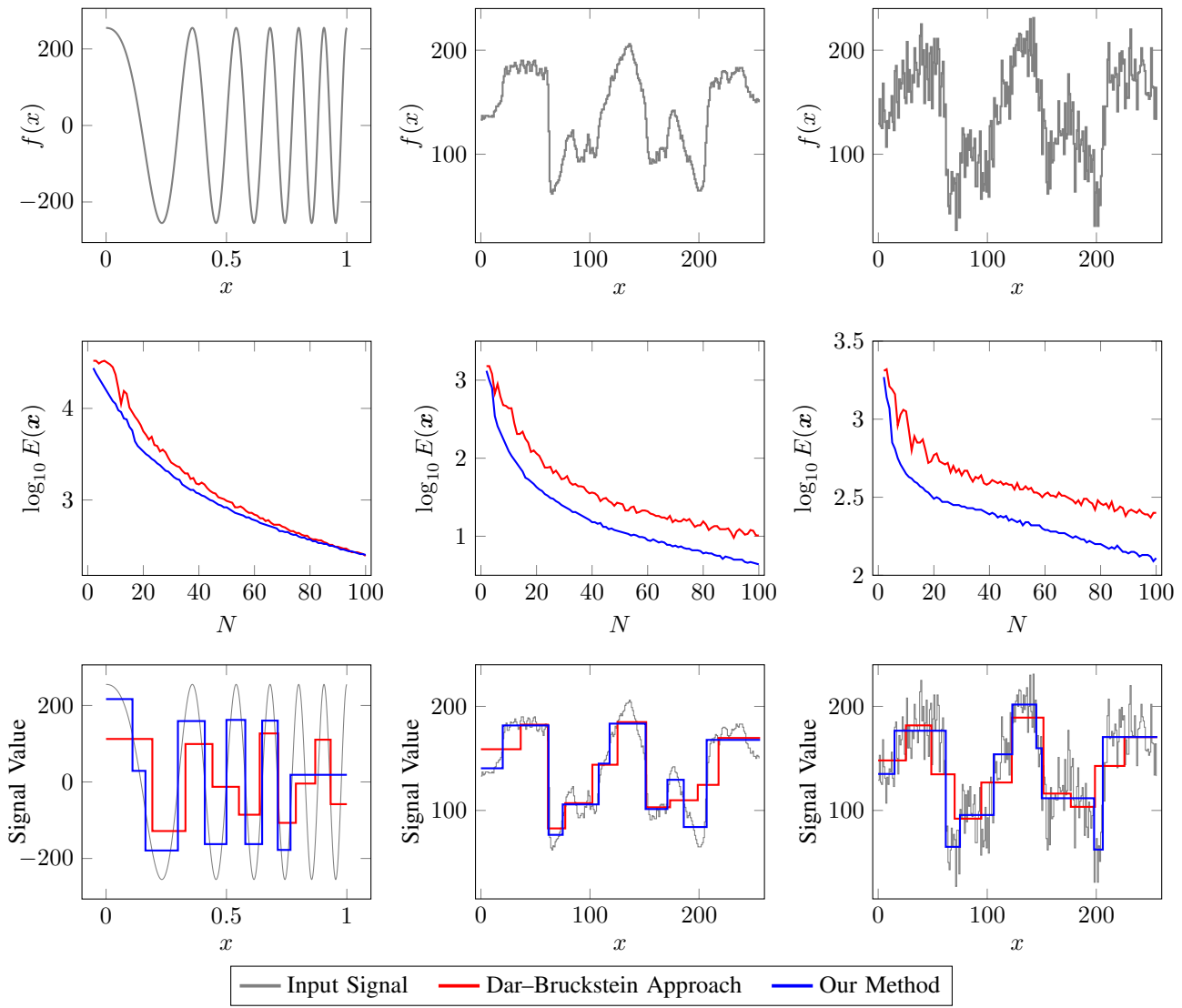
\begin{figure*}
\centering
\begin{tikzpicture}
\matrix {
    \begin{axis}[%
        domain=0:1,
        xlabel={$x$},
        ylabel={$f(x)$},
        y label style={at={(axis description cs:-0.1,.5)}},
        width=0.32\textwidth]
    \addplot[samples=1000, gray, thick]
        {255*cos(deg(2*pi*x*(1+5*x)))};
    \end{axis}
    &
    \begin{axis}[%
        xmin=-5, xmax=260,
        ymin=15, ymax=240,
        xlabel={$x$},
        ylabel={$f(x)$},
        y label style={at={(axis description cs:-0.15,.5)}},
        width=0.32\textwidth]
    \addplot[gray, const plot, thick] table [x={x}, y={fx}]
        {data/img_sig.dat};
    \end{axis}
    &
    \begin{axis}[%
        xmin=-5, xmax=260,
        ymin=15, ymax=240,
        xlabel={$x$},
        ylabel={$f(x)$},
        y label style={at={(axis description cs:-0.15,.5)}},
        width=0.32\textwidth]
    \addplot[gray, const plot, thick] table [x={x}, y={fx}]
        {data/img_sig_noise.dat};
    \end{axis}
    \\[1em]
    \begin{axis}[%
        xmin=-2, xmax=102,
        xlabel={$N$},
        ylabel={$\log_{10} E(\bm{x})$},
        width=0.32\textwidth]
    \addplot[red, thick] table [x={N}, y={log10E}]
        {data/chirp_mse_dar.dat};
    \addplot[blue, thick] table [x={N}, y={log10Min}]
        {data/chirp_mse_pso.dat};
    \end{axis}
    &
    \begin{axis}[%
        xmin=-2, xmax=102,
        ymin=0.5, ymax=3.5,
        xlabel={$N$},
        ylabel={$\log_{10} E(\bm{x})$},
        width=0.32\textwidth]
    \addplot[red, thick] table [x={N}, y={log10E}]
        {data/img_sig_mse_dar.dat};
    \addplot[blue, thick] table [x={N}, y={log10Min}] 
        {data/img_sig_mse_pso.dat};
    \end{axis}
    &
    \begin{axis}[%
        xmin=-2, xmax=102,
        ymin=2.0, ymax=3.5,
        xlabel={$N$},
        ylabel={$\log_{10} E(\bm{x})$},
        legend cell align=left,
        width=0.32\textwidth]
    \addplot[red, thick] table [x={N}, y={log10E}]
        {data/img_sig_noise_mse_dar.dat};
    \addplot[blue, thick] table [x={N}, y={log10Min}] 
        {data/img_sig_noise_mse_pso.dat};
    \end{axis}
    \\[1em]
    \begin{axis}[%
        domain=0:1,
        xlabel={$x$},
        ylabel={Signal Value},
        y label style={at={(axis description cs:-0.15,.5)}},
        width=0.32\textwidth,
        legend columns={-1},
        legend to name={legend},
        legend image post style={line width=0.5mm},
        legend entries={{Input Signal}, {Dar--Bruckstein Approach}, {Our 
        Method}},
        legend style={/tikz/every even column/.append style={column sep=1em}}]
    \addplot[samples=1000, gray]
        {255*cos(deg(2*pi*x*(1+5*x)))};
    \addplot[red, const plot, thick] table [x={x}, y={fx}] 
        {data/chirp_approx_10_dar.dat};
    \addplot[blue, const plot, thick] table [x={x}, y={fx}] 
        {data/chirp_approx_10_pso.dat};
    \end{axis}
    &
    \begin{axis}[%
        xmin=-5, xmax=260,
        ymin=15, ymax=240,
        xlabel={$x$},
        ylabel={Signal Value},
        y label style={at={(axis description cs:-0.15,.5)}},
        width=0.32\textwidth]
    \addplot[gray, const plot]
        table [x={x}, y={fx}] {data/img_sig.dat};
    \addplot[red, const plot, thick] table [x={x}, y={fx}] 
        {data/img_sig_approx_10_dar.dat};
    \addplot[blue, const plot, thick] table [x={x}, y={fx}] 
        {data/img_sig_approx_10_pso.dat};
    \end{axis}
    &
    \begin{axis}[%
        xmin=-5, xmax=260,
        ymin=15, ymax=240,
        xlabel={$x$},
        ylabel={Signal Value},
        y label style={at={(axis description cs:-0.15,.5)}},
        width=0.32\textwidth]
    \addplot[gray, const plot]
        table [x={x}, y={fx}] {data/img_sig_noise.dat};
    \addplot[red, const plot, thick] table [x={x}, y={fx}] 
        {data/img_sig_noise_approx_10_dar.dat};
    \addplot[blue, const plot, thick] table [x={x}, y={fx}] 
        {data/img_sig_noise_approx_10_pso.dat};
    \end{axis}
    \\
};
\end{tikzpicture}
\ref{legend}
\caption{
{\bf Left column:} Results for the chirp signal.
{\bf Central column:} Results for the image signal.
{\bf Right column:} Results for the image signal degraded by additive 
                    Gaussian noise.
% with $\mu = 0$ and $\sigma = 20$.
{\bf Top row:} Original signals.
{\bf Middle row:} MSE for the Dar--Bruckstein approach and 
                  $\min_{\mbox{\tiny{MSE}}}$ for our method as a function of 
                  the sample number $N$.
{\bf Bottom row:} Signal approximations for $10$ samples.
We present the results with lowest MSE from all $50$ runs of the PSO algorithm.}
\label{fig:signals}
\end{figure*}

We have implemented the Dar--Bruckstein model as is proposed in 
\cite[Subsection II. A.]{DB16} using \eqref{eq:yehuda_sol}. For the PSO 
algorithm for optimising our direct model, we adhere to \cite{ZBCR13}.
As PSO parameters we use a maximum number of $10000$ iterations, a swarm 
size of $n = 1000$, and a neighbourhood size of $K = 20$. We reset the 
neighbourhood structures after $15$ iterations with no change in the 
global minimum. Following \cite{ZBCR13}, we choose 
$c_1 = c_2 = 0.5 + \ln(2)$ and
$\omega = 1 / (2 \ln(2))$.
Since the PSO algorithm involves randomisation, the quality of 
multiple program runs with identical parameters may differ somewhat. 
Thus, for every sample number $N$, we run the PSO algorithm 
$50$ times and report the mean $\mu_{\mbox{\tiny{MSE}}}$, the standard 
deviation $\sigma_{\mbox{\tiny{MSE}}}$, the minimum $\min_{\mbox{\tiny{MSE}}}$, 
and the maximum $\max_{\mbox{\tiny{MSE}}}$ of our MSE computations.
%
% of the MSE.
%Knowing these 
%values allows to improve the quality by choosing the best result of multiple
%program runs: Probability theory implies that one can expect to find one 
%MSE result below $ \mu_{\mbox{\tiny{MSE}}} - \sigma_{\mbox{\tiny{MSE}}}$ 
%in $1$ out of $7$ runs, and one MSE below $\mu_{\mbox{\tiny{MSE}}} - 
%2 \sigma_{\mbox{\tiny{MSE}}}$ happens in about $1$ of $44$ runs.
%
%We denote the lowest and highest MSE of all $50$ runs by 
%$\min_{\mbox{\tiny{MSE}}}$ and $\max_{\mbox{\tiny{MSE}}}$.
% 
These MSE results are listed in Tab.~\ref{tab:results}, and 
Fig.~\ref{fig:signals} displays the signals, error graphs, and 
approximations.   

In our first experiment we consider the chirp signal 
$f(x) = 255 \, \cos(2 \pi x (1 + 5x))$ within the interval $[0,1]$; 
see Fig.~\ref{fig:signals}, top left. It was also studied in \cite{DB16}.
It constitutes a prototype for a smooth signal, which is nevertheless
challenging in its high frequent right part.
We observe that for $N \le 50$, the Dar--Bruckstein approach 
performs consistently worse than the direct optimisation method.
It appears to suffer from violations of the local linearity assumption.
For larger values of $N$, this effect vanishes more and more, and 
the Dar--Bruckstein algorithm reaches a comparable quality as the 
direct approach. It is even slighly ahead for very large $N$. This 
is caused by the nonoptimality of the PSO algorithm: Its performance 
deteriorates somewhat for the more complex optimisation problems that 
arise for large $N$. We expect that this can be solved with more advanced 
optimisation techniques that we will consider in our future research. 

In our second experiment, we use the nonsmooth real-world signal shown
%in Fig.~\ref{fig:signals}, top right.
in Fig.~\ref{fig:signals}, top centre.
It represents line 51 of the 8-bit 
test image {\em trui}, which depicts a lady with a scarf. 
Within each of its $256$ pixels, the function values are regarded constant.
Having many jump discontinuities, this signal violates both the 
smoothness and the local linearity assumption. Thus, it is not surprising
that it poses substantial challenges for the Dar--Bruckstein model: We 
observe that the Dar--Bruckstein MSE is about twice as large as the
MSE of the direct optimisation method, and that this factor remains
also for large values of $N$. The direct optimisation approach clearly 
benefits from its absence of any smoothness or local linearity requirements.  

For our third experiment, we degrade the previous image signal  
by additive Gaussian noise with zero mean and standard deviation 
$20$; see Fig.~\ref{fig:signals}, top right. This takes the violation 
of the smoothness and local linearity assumption to the extremes.
While the approximation quality of both approaches deteriorates in
comparison to the noise-free scenario, the MSE of the Dar--Bruckstein 
method remains almost twice as large as the one of our direct optimisation
approach. 

Without going into details, we remark that in all experiments and for
both methods, we found substantial deviations from perfect error balancing. 
% For a fixed $N$, averaged standard deviations of the error $e_i$ for a method
% and a specific signal amounted to $0.25$ to $1.1$ times their mean 
% value.
% This effect was least pronounced when applying the Dar--Bruckstein method 
% with large $N$ to the smooth chirp signal.
For the Dar--Bruckstein method this shows that the local linearity assumption 
is still not fully met. For our direct optimisation approach it indicates that 
the optimal solution does not satisfy ideal error balancing. This is in 
agreement with the counterexample in Fig.~\ref{fig:example_signal}(a).

%...........................................................................

%...........................................................................

%
\begin{table}
\caption{Approximation Quality of the Dar--Bruckstein (DB) Approach and 
Our Direct Optimisation Method}
\begin{center}
\begin{tabular}{|c|r|r|r|r|r|r|}
\hline
&
\multicolumn{1}{|c|}{\multirow{2}{*}{\textbf{N}}} &
\multicolumn{1}{|c|}{\textbf{DB}} &
\multicolumn{4}{|c|}{\textbf{Our Method}}\\
\cline{3-7}
&
&
\multicolumn{1}{|c|}{MSE} &
\multicolumn{1}{|c|}{$\mu_{\mbox{\tiny{MSE}}}$} &
\multicolumn{1}{|c|}{$\sigma_{\mbox{\tiny{MSE}}}$} &
\multicolumn{1}{|c|}{$\min_{\mbox{\tiny{MSE}}}$} &
\multicolumn{1}{|c|}{$\max_{\mbox{\tiny{MSE}}}$}\\
\hline
\multirow{11}{*}{\rotatebox{90}{Chirp Signal}} &
5   & 32166.88 & 19055.49 &   0.00 & 19055.49 & 19055.49\\
\cline{2-7}
& 10  & 23655.94 & 12014.77 & 459.47 & 11131.75 & 12431.60\\
\cline{2-7}
& 20  &  5661.01 &  4556.37 & 632.89 &  3403.23 &  5998.27\\
\cline{2-7}
& 30  &  2590.87 &  2037.90 &  67.90 &  1906.95 &  2285.76\\
\cline{2-7}
& 40  &  1477.48 &  1218.49 &  37.17 &  1177.55 &  1357.26\\
\cline{2-7}
& 50  &   975.93 &   854.12 &  20.76 &   823.13 &   907.48\\
\cline{2-7}
& 60  &   686.30 &   624.44 &  13.37 &   601.65 &   665.50\\
\cline{2-7}
& 70  &   510.96 &   479.90 &   9.26 &   460.10 &   513.00\\
\cline{2-7}
& 80  &   377.20 &   383.08 &   6.95 &   367.62 &   400.07\\
\cline{2-7}
& 90  &   307.64 &   311.04 &   5.11 &   301.63 &   328.83\\
\cline{2-7}
& 100 &   247.84 &   258.05 &   4.25 &   249.73 &   269.60\\
\hline
\multicolumn{7}{c}{}\\[-1ex]
\hline
\multirow{11}{*}{\rotatebox{90}{Image Signal}} &
5   & 674.31 &  347.36 &  0.00 &  347.36 & 347.36\\
\cline{2-7}
& 10  & 436.22 &  126.37 &  3.51 &  123.47 & 130.96\\
\cline{2-7}
& 20  & 115.28 &   47.83 &  3.75 &   43.05 &  60.15\\
\cline{2-7}
& 30  &  54.46 &   26.28 &  1.28 &   24.03 &  29.48\\
\cline{2-7}
& 40  &  36.58 &   17.39 &  0.79 &   15.15 &  18.83\\
\cline{2-7}
& 50  &  26.34 &   12.63 &  0.60 &   11.41 &  14.28\\
\cline{2-7}
& 60  &  20.90 &   10.23 &  0.42 &    8.88 &  11.20\\
\cline{2-7}
& 70  &  16.25 &    8.34 &  0.40 &    7.54 &   9.28\\
\cline{2-7}
& 80  &  15.47 &    7.01 &  0.43 &    6.09 &   8.29\\
\cline{2-7}
& 90  &  11.13 &    5.87 &  0.31 &    5.21 &   6.67\\
\cline{2-7}
& 100 &  10.20 &    5.09 &  0.28 &    4.33 &   5.64\\
\hline
\multicolumn{7}{c}{}\\[-1ex]
\hline
\multirow{11}{*}{\rotatebox{90}{Noisy Image Signal}} &
5   & 1532.52 & 709.29 & 0.00 & 709.29  & 709.29\\
\cline{2-7}
& 10  & 1130.94 & 454.99 & 6.54 & 451.57  & 474.21\\
\cline{2-7}
& 20  &  588.69 & 332.17 & 8.15 & 312.44  & 350.30\\
\cline{2-7}
& 30  &  435.49 & 286.25 & 4.90 & 277.01  & 300.15\\
\cline{2-7}
& 40  &  391.56 & 262.20 & 5.33 & 244.90  & 273.72\\
\cline{2-7}
& 50  &  391.27 & 242.26 & 8.45 & 219.93  & 262.93\\
\cline{2-7}
& 60  &  330.87 & 212.69 & 7.68 & 196.22  & 229.98\\
\cline{2-7}
& 70  &  301.52 & 193.08 & 8.39 & 174.53  & 212.21\\
\cline{2-7}
& 80  &  288.86 & 170.87 & 6.63 & 157.66  & 189.61\\
\cline{2-7}
& 90  &  258.99 & 156.27 & 7.25 & 139.10  & 172.14\\
\cline{2-7}
& 100 &  250.82 & 142.21 & 6.97 & 128.11  & 164.89\\
\hline
\end{tabular}
\label{tab:results}
\end{center}
\end{table}
%

%%%%%%%%%%%%%%%%%%%%%%%%%%%%%%%%%%%%%%%%%%%%%%%%%%%%%%%%%%%%%%%%%%%%%%%

\section{Conclusions and Outlook}
\label{sec:summary}

We have studied piecewise constant signal approximations that possess 
a specified number of samples and aim at minimising the MSE to the 
original signal. 

As a first contibution, we have provided a simple alternative derivation 
of the recent Dar--Bruckstein approach. It enabled us to analyse 
the limitations of the method in detail. We have shown that the 
quality of the Dar--Bruckstein approach is compromised, if smoothness 
or local linearity are violated, or if error balancing leads to 
suboptimal solutions. 

In a second step, these insights have triggered us to consider a 
direct optimisation model that renounces all three 
assumptions. It gives a globally optimal solution if one can 
solve the corresponding nonconvex optimisation problem exactly. 
Our experiments show that already a suboptimal particle swarm 
optimisation algorithm yields good local minima in practice.

Evaluating the quality of both approaches for a smooth, a nonsmooth,
and a noisy signal has demonstrated that the direct method 
offers better quality if the signal lacks smoothness or the number 
of samples is low. For smooth signals with many samples, the 
Dar--Bruckstein algorithm remains a simple and efficient alternative 
with comparable quality.

In our ongoing work, we are investigating alternative optimisation 
approaches for the nonconvex model. Moreover, we will extend our
research on piecewise constant approximations to the multidimensional 
case, and we are going to study suitable applications, e.g.~in data 
compression and quantisation.

Apart from its specific contibutions to signal approximation, our 
paper may also be of more general relevance by disproving the 
optimality of the principle of error balancing for optimisation
problems that involve the $\ell^2$-norm. This principle is 
omnipresent in numerous applications, ranging from scientific 
computing to computer graphics. 
Thus, understanding its suboptimality in more detail may also pave 
the road to better algorithms beyond the field of signal processing.

%%%%%%%%%%%%%%%%%%%%%%%%%%%%%%%%%%%%%%%%%%%%%%%%%%%%%%%%%%%%%%%%%%%%%%%

\section{Acknowledgement}

We thank Tobias Alt and Matthias Augustin for valuable comments and
fruitful discussions.

%%%%%%%%%%%%%%%%%%%%%%%%%%%%%%%%%%%%%%%%%%%%%%%%%%%%%%%%%%%%%%%%%%%%%%%

\bibliographystyle{IEEEtran}
\bibliography{IEEEabrv,bergerhoff.bib}

\end{document}